\documentclass[conference,compsoc]{IEEEtran}
\IEEEoverridecommandlockouts

\usepackage{microtype}
\usepackage{graphicx}
\usepackage{subfigure}
\usepackage{booktabs} 

\usepackage{hyperref}
\usepackage{url}
\usepackage{algorithmic}
\usepackage{textcomp}
\usepackage{xurl}
\usepackage[table,dvipsnames]{xcolor}
\usepackage{multirow}
\usepackage{rotating}
\usepackage{enumitem}
\usepackage{wrapfig}
\usepackage{bmpsize}
\usepackage[]{footmisc}
\usepackage{caption}
\usepackage{placeins}
\usepackage{colortbl} 
\usepackage{subcaption}
\usepackage[most]{tcolorbox}

\usepackage{cleveref}
\usepackage{comment}

\usepackage[color=yellow]{todonotes}

\newcommand{\at}[1]{\textit{$T_{#1}$}}
\newcommand{\prop}[1]{\textit{$P_{#1}$}}
\newcommand{\sys}{\textit{Cascade }}

\newcommand{\colorrow}[2]{%
  \gdef\do##1{\cellcolor{#2}##1\ignorespaces}%
  \expandafter\do#1\unskip%
}

\newcommand{\circnum}[1]{%
  \tikz[baseline=(char.base)]{
    \node[shape=circle,fill=black,inner sep=1pt] (char)
    {\color{white}\small #1};}}

\setlist{noitemsep, leftmargin=*, topsep=0pt, partopsep=0pt}

\def\BibTeX{{\rm B\kern-.05em{\sc i\kern-.025em b}\kern-.08em
    T\kern-.1667em\lower.7ex\hbox{E}\kern-.125emX}}

\begin{document}

\title{Cascade: Composing Software-Hardware Attack Gadgets for Adversarial Threat Amplification in Compound AI Systems}

\author{\IEEEauthorblockN{ 
Sarbartha Banerjee\IEEEauthorrefmark{1}\IEEEauthorrefmark{2}\IEEEauthorrefmark{6},\;
Prateek Sahu\IEEEauthorrefmark{1}\IEEEauthorrefmark{2},\;
Anjo Vahldiek-Oberwagner\IEEEauthorrefmark{3},\;
Jose Sanchez Vicarte\IEEEauthorrefmark{5},\;
Mohit Tiwari\IEEEauthorrefmark{2}\IEEEauthorrefmark{4}} \vspace{0.3em}
\IEEEauthorblockA{\IEEEauthorrefmark{2}The University of Texas at Austin \;\; \IEEEauthorrefmark{3}Intel Labs \;\; \IEEEauthorrefmark{4}Symmetry Systems \;\; \IEEEauthorrefmark{5}Microsoft \;\; \IEEEauthorrefmark{6}Georgia Tech}
}

\maketitle
\thispagestyle{plain}

\begingroup\begin{NoHyper}\renewcommand\thefootnote{\IEEEauthorrefmark{1}}
\footnotetext{Sarbartha Banerjee and Prateek Sahu are equal contributors.}\end{NoHyper}

\vskip 0.3in

\begin{abstract}
Rapid progress in generative AI has given rise to \textit{Compound AI systems} - pipelines comprised of multiple large language models (LLM), software tools and database systems.
Compound AI systems are constructed on a layered traditional software stack running on a distributed hardware infrastructure.
Many of the diverse software components are vulnerable to traditional security flaws documented in the Common Vulnerabilities and Exposures (CVE) database, while the underlying distributed hardware infrastructure remains exposed to timing attacks, bit-flip faults, and power-based side channels.
Today, research targets LLM-specific risks like model extraction, training data leakage, and unsafe generation -- overlooking the impact of traditional system vulnerabilities.

This work investigates how traditional software and hardware vulnerabilities can complement LLM-specific algorithmic attacks to compromise the integrity of a compound AI pipeline.
We demonstrate two novel attacks that combine system-level vulnerabilities with algorithmic weaknesses:
\textbf{(1)} Exploiting a software code injection flaw along with a guardrail Rowhammer attack to inject an unaltered jailbreak prompt into an LLM, resulting in an AI safety violation, and
\textbf{(2)} Manipulating a knowledge database to redirect an LLM agent to transmit sensitive user data to a malicious application, thus breaching confidentiality.
These attacks highlight the need to address traditional vulnerabilities; we systematize the attack primitives and analyze their composition by grouping vulnerabilities by their objective and mapping them to distinct stages of an attack lifecycle.
This approach enables a rigorous red-teaming exercise and lays the groundwork for future defense strategies.

\end{abstract}




\section{Introduction}
Large language models (LLMs) are rapidly reshaping industries -- from language translation and art to finance and engineering -- by offering unprecedented capabilities in natural language processing and generation. 
Recent systems such as GPT-4o \cite{ChatGPT}, Gemini \cite{Gemini}, Deepseek \cite{deepseek} and Microsoft Copilot \cite{Copilot} comprise multiple specialized LLMs - each fine-tuned for domain-specific expertise - alongside a contextual knowledge database, software tools for task execution, and guardrails to enforce response safety and correctness.
These compound AI systems~\cite{compound-ai-blog} are used in a wide range of applications, including conversational agents~\cite{ChatGPT,Gemini}, software development~\cite{gitCopilot}, productivity tools~\cite{Copilot}, content creation, robotic agents and legal compliance.

\begin{figure*}[t!]
    \centering
    \includegraphics[width=1\linewidth]{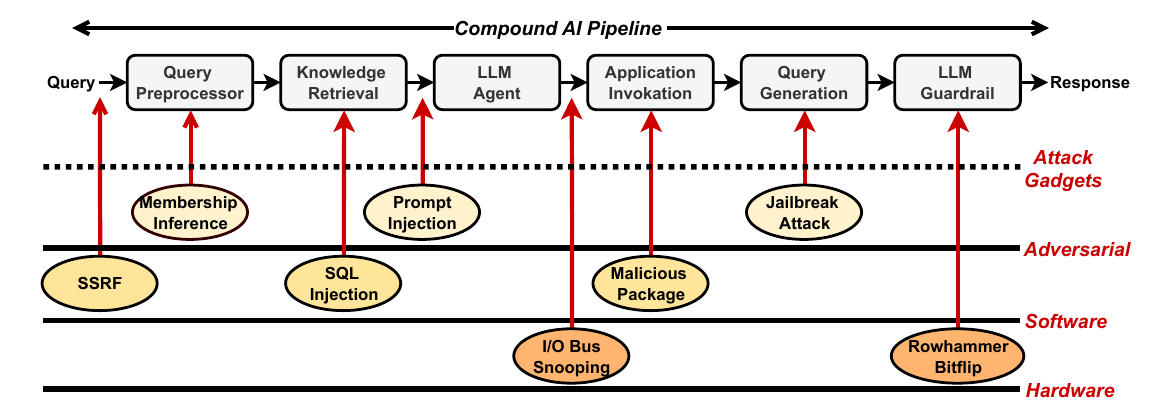}
    \vspace{-2em}
    \caption{\bf The building blocks of a Compound AI pipeline with cross-stack attack gadgets comprising of adversarial attacks, software vulnerabilities and hardware side-channels.
    }
    \label{fig:AI-Stack-intro}
\end{figure*}

~\Cref{fig:AI-Stack-intro} provides a schematic of the building blocks of a Compound AI pipeline.
The user interacts with the system via a web or chatbot interface. 
Each query is first handled by a preprocessor that interprets the context and transforms it into an enriched request.
Next, the knowledge retrieval module searches for relevant information from a database or the web and appends it to the enriched request, which is then sent to an LLM agent model.
The agent decomposes the user query and orchestrates the appropriate applications to handle each component of the request. These applications may include software tools such as code interpreters, presentation tools, or lambda-based microservices that perform tasks such as retrieving weather data, adding calendar events, or executing calculations. The resulting outputs are then aggregated into a coherent response by a query generation LLM. Finally, this response is evaluated by a guardrail LLM to ensure accuracy and safety before being delivered to the user.
Underlying the application layer, the pipeline comprises a diverse set of software components,
including LLM frameworks (e.g., LangChain \cite{Langchain}, Ollama \cite{ollama}), data structure stores (e.g., Redis \cite{redis}, Data Lakes \cite{datalakes}, MySQL \cite{mysql}), software utilities (e.g., Java APIs, Node.js, Python FastAPI), foundational packages (e.g., PyTorch \cite{Pytorch}, TensorFlow \cite{Tensorflow},  Apache Spark \cite{Spark}, Kubernetes \cite{Kubernates}), and low-level libraries (e.g., cuDNN \cite{Cudnn}, OpenBLAS \cite{openblas}, OneAPI \cite{oneapi}).
It is deployed on a distributed hardware infrastructure that spans multiple compute nodes (CPUs, GPUs, specialized accelerators), memory modules (DRAM, HBM), interconnects (NVLink, PCIe, Mellanox), and storage devices (SSDs, NVRAM).

As Compound AI pipelines increasingly process sensitive user data, such as emails, photos, and medical records, and are deployed in critical domains like autonomous vehicles, social media platforms, and warehouse automation, they become prime targets for adversarial attacks that threaten the safety, confidentiality, and integrity of these systems.
Adversarial attacks target the LLM model algorithm to violate training data privacy (ex., Membership inference attacks \cite{shokri2017membership}), tamper training data (ex., Data poisoning attacks \cite{tolpegin2020data}), extract LLM parameters (ex., Model stealing attacks \cite{hua2018reverse,tramer2016stealing}) or tamper response safety (ex. Jailbreak attacks \cite{liu2023autodan}). 
While significant research has focused on adversarial attacks against AI models, the role of software vulnerabilities -- memory safety, code injection, buffer overflow -- and hardware side-channels such as timing leaks, bit-flip faults, and power analysis, remains largely neglected.
However, the growing complexity of AI pipelines amplifies the potential for system-level attacks to complement adversarial techniques, enabling more effective exploitation.
Firstly, AI pipelines increasingly incorporate non-AI components -- such as knowledge databases, applications, and orchestration frameworks -- that are primarily susceptible to system-level attacks.
Secondly, downstream models are often invoked indirectly through tool usage or constrained by input/output filtering mechanisms, limiting direct interaction. While algorithmic attacks can still achieve some success with indirect model access~\cite{bullwinkelLessonsRedTeaming2025}, system-level attack gadgets can bypass these indirections, amplifying the overall effectiveness of the attack.
For instance, software vulnerabilities like server-side request forgery can leak user query, SQL injection can tamper the knowledge database and a malicious package can serve as a backdoor as shown in~\cref{fig:AI-Stack-intro}.
Similarly, hardware attacks like I/O bus snooping can fingerprint agent decisions from inter-component data transfers or a rowhammer bitflip \cite{kim2014flipping} can alter guardrail safety decision.
Third, system-level attacks are inherently more difficult to mitigate, as their underlying vulnerabilities lie outside the scope of algorithmic defenses. For example, a guardrail bit-flip attack operates independently of the model architecture and can persist across retraining. Effectively detecting such exploits requires defenders to look beyond model logs and adopt a holistic view of the entire software and hardware stack.

In this paper, we investigate software and hardware attack gadgets within a Compound AI pipeline and demonstrate how cross-stack composition of these gadgets can be leveraged to exploit the AI inference process.
We demonstrate a novel end-to-end attack that violates AI safety by jailbreaking an LLM model in the presence of a query enhancer and a guardrail LLM. 
The query enhancer is bypassed via a code injection vulnerability, while the guardrail is circumvented using a Rowhammer-based fault injection.
This attack illustrates how system-level gadgets can grant an attacker direct query access to a downstream LLM, bypassing intermediate controls.
We also identify alternative gadget compositions capable of achieving similar attack goals, laying a foundation for future red-teaming efforts and informing defense strategy decisions for Compound AI systems.

In summary, the key contributions of the paper are:
\begin{enumerate}[leftmargin=*]
    \item We curated a corpus of hundreds of attack gadgets spanning algorithmic, software, and hardware layers to investigate how system-level vulnerabilities complement and amplify adversarial threats in compound AI systems.
    \item We present the \sys Red Teaming Framework, which generates end-to-end attack chains by mapping an adversary’s goals and capabilities to a curated set of algorithmic, software, and hardware attack gadgets targeting multiple AI pipeline components.
    \item We demonstrate the usefulness of the cascade framework with several cross-layer attack gadget compositions, including a concrete attack violating ai safety even with the presence of pipeline protections like AI guardrails.
\end{enumerate}

\section{Security of Compound AI Systems}

An AI system features a layered architecture, with the \textit{application layer} encompassing multiple trained LLMs, vector databases, LLM-driven applications, and AI agents. 
Below it, the \textit{software layer} comprises of frameworks, packages, and libraries that supports these application components. This includes pipeline-building frameworks like LangChain, training and fine-tuning libraries such as PyTorch and TensorFlow, database backends like Apache Spark, MongoDB, and Redis, as well as programming environments for developing LLM applications and utilities. Libraries, device-drivers and other low-level dependencies also form part of this layer.
This comprehensive software layer operates on a distributed hardware backend. Multiple GPUs are employed to execute LLMs, while storage devices such as SSDs support vector databases. High-bandwidth interconnects -- both local (e.g., PCIe, NVLink) and remote (e.g., InfiniBand) -- enable efficient data transfer across the system.

\subsection{Components in Compound AI Pipeline}
As we move from simple LLM deployments to full-fledged production applications powered by LLMs, the challenge of securing them increases significantly. Safety of responses and security of training data drove industry to retrain models with in-built safety categories as well as research into diffusion model architectures that are not susceptible to memorization~\cite{memorization} or hallucinations. Productivity and the high cost of retraining drove towards more engineering solutions for safeguarding LLM pipelines:

\textbf{Guardrails:} To address the growing concern around prompt injection attacks and use of jailbreak prompts to elicit unsafe responses, guardrail models were designed. Guardrail models~\cite{rebedea2023nemoguardrailstoolkitcontrollable,AWSBedrock,guardrailai} are trained on large categories of information that can be deemed unsafe, unethical or harmful. In addition, such models also support few-shot learning for newer categories as per developer's requirements. While LLMs themselves, the output of guardrails are binary safe/unsafe responses to the query. The usage of guardrail models have been shown to be effective against prompt injection attacks where the query tries to confuse the language model into providing harmful content by providing malicious instructions~\cite{DAN}.

\textbf{Query enhancers:} Query enhancers mitigate adversarial inputs by rewriting prompts by removing irrelevant prefixes and suffixes, instead of making binary allow/block decisions like guardrails. Modern AI pipelines adopt diverse sanitization methods such as perplexity-based filtering~\cite{alon2023detectinglanguagemodelattacks}, re-tokenization~\cite{cao2024defendingalignmentbreakingattacksrobustly}, paraphrasing~\cite{jain2023baselinedefensesadversarialattacks}, and randomized smoothing (e.g., SmoothLLM~\cite{robey2023smoothllm}) to enhance response robustness. These transformations complicate adversarial exploration since jailbreaks depend on precise keywords or token sequences, making simple suffix or prefix attacks ineffective. 

\textbf{Grounding:} Modern compound AI systems rely on accurate knowledge databases, where each entry is verified through grounding by human administrators~\cite{Lamini,Predibase} or LLMs~\cite{promptshield,JinaGrounding,GeminiFC}. Grounding modules cross-reference external sources, resolve conflicts, and block low-perplexity outputs to prevent hallucinations and poisoning attacks like PoisonedRAG~\cite{poisonedrag}. Access control mechanisms and IAM integrations (e.g., M365 Copilot) restrict unauthorized access to sensitive services and documents. These defenses mitigate threats such as ConfusedPilot~\cite{confusedpilot}, which exploit misconfigured privileges to spread misinformation.


\subsection{Cross-stack attack gadgets}
While adversarial attacks primarily target the \textit{application layer}, 
the vulnerabilities in \textit{software} and \textit{hardware} layers can complement such attacks.
Growing application-layer defenses have made standalone adversarial attacks harder, so attackers increasingly combine application, software, and hardware flaws into multi-step chains. Distributed backends let adversaries target isolated components (e.g., use code injection to crash a pipeline node stealthily) and then exploit those failures to weaken overall protections. Practical cross-layer primitives include SQL injection against vector stores, malicious third-party packages that exfiltrate sensitive data, man-in-the-middle tampering of inter-component traffic, and privilege-escalation–driven resource exhaustion. Hardware weaknesses such as cache timing, memory bitflips, I/O snooping, and storage attacks can similarly lower the bar for successful attacks\cite{yarom2014flush+,kim2014flipping,lee2020off,xiao2015mitigating}. With cross-stack attack vectors, adversaries can exploit software and hardware flaws in compound AI pipelines to amplify algorithmic attacks and evade detection.

Securing the software and hardware stack in compound AI systems is exceptionally difficult due to their scale, dependency complexity, and heterogeneous infrastructure. Frequent version mismatches, limited integrity protection, and persistent side-channel risks make remediation slow and impractical, allowing system vulnerabilities to outlast and outweigh algorithmic ones.


\section{Attack Gadget Systematization}
\label{sec:attack_gadget_characterization}

\begin{figure}[t!]
    \centering
    \includegraphics[width=1.05\linewidth]
    {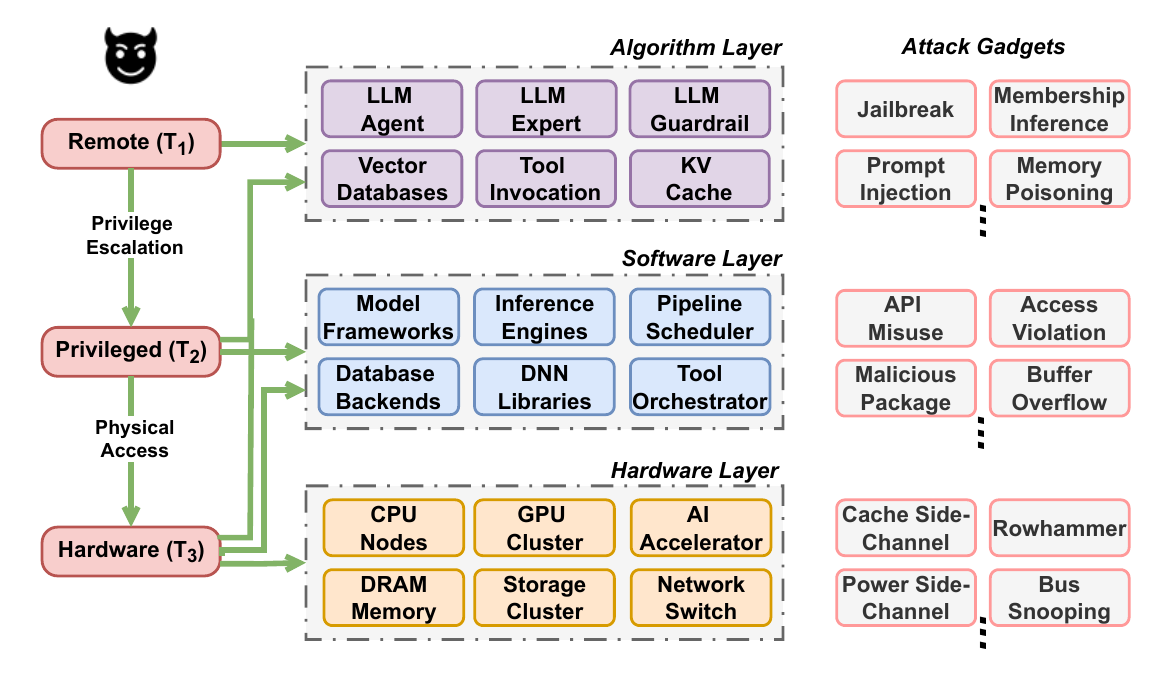}
    \vspace{-2em}
    \caption{\bf The building blocks of a Compound AI pipeline with cross-stack attack gadgets comprising of adversarial attacks, software vulnerabilities and hardware side-channels.
    }
    \label{fig:Threat-Model-Characterization}
\end{figure}

Attack gadgets span across multiple stack layers for each component in a Compound AI pipeline. 
The \sys framework takes the 
\textbf{(1)} the deployed AI pipeline, 
\textbf{(2)} the attacker goal, and 
\textbf{(3)} the attacker capability 
to shortlist the available attack gadgets from a repository of cross-attack vectors. 
Next, the \sys framework either finds a single attack gadget  or
compose multiple gadgets to achieve the attacker's goal. 
The purpose of the \sys red-teaming framework is to navigate the vast cross-stack cross-component attack surface
and find possible attack compositions that help system designers to find unexpected system vulnerabilities in Compound AI systems.

\subsection{Security properties}
\label{subsec:sec_property}
The \sys red-teaming framework takes the attacker's goal as an input and maps it to the violation of one of the many security properties. 
We enlist the following security properties that serve as the outcome of a successful attack:

\noindent\textbf{{Confidentiality Property (\prop{1})}}
The confidentiality property ensures that the AI pipeline does not leak any secret asset classified by the owner of each pipeline component. 
This includes the following and is not limited to:
\textbf{(1)} The privacy of the LLM training data, 
\textbf{(2)} The model parameters and hyperparameters of proprietary models,
\textbf{(3)} The composition of an AI pipeline, 
\textbf{(4)} Secret and access controlled entry in vector databases,
\textbf{(5)} Privileged data structures and scheduling mechanisms of the deployment platform.

\noindent\textbf{{Integrity Property (\prop{2})}}
The integrity property ensures that an attacker does not alter each component and the connection between them.
This includes the following and is not limited to:
\textbf{(1)} Tampering LLM accuracy by modifying model parameters or poisoning training data,
\textbf{(2)} Tampering the query tokens, intermediate query context or memory,
\textbf{(3)} Inserting malicious data or tampering the knowledge in RAG databases,
\textbf{(4)} Inserting malicious packages or tools in the agent tool repository.

\noindent\textbf{{Safety Property (\prop{3})}}
The safety property ensures that an attacker does not generate harmful or incorrect content.
This includes the following and is not limited to: 
\textbf{(1)} The pipeline generates illicit, harmful or abusive output,
\textbf{(2)} AI code generation pipeline outputs incorrect or vulnerable code,
\textbf{(3)} The pipeline output is incorrect or has low confidence.

\noindent\textbf{{Availability Property (\prop{4})}}
The availability property ensures that an attacker does not interfare with the pipeline execution. 
This includes the following and is not limited to:
\textbf{(1)} Crashing or unavailability of pipeline block, 
\textbf{(2)} Tampering with the deployment scheduler to delay or omit a pipeline block,
\textbf{(3)} Unauthorized use of system resources leading to resource exhaustion.
\textbf{(4)} Replacement of LLM models or tools for generation of an inferior output.

\noindent\textbf{{Authorization Property (\prop{5})}}
The authorization property ensures that an attacker does not get unwarranted access to the AI pipeline.
This includes the following and is not limited to:
\textbf{(1)} Gaining access to specific pipeline components through crafted queries,
\textbf{(2)} Admin access to the deployed hypervisor or hardware beyond the designated sandbox,
\textbf{(3)} Unauthorized access to the vector database, 
\textbf{(4)} Add, modify or delete system files caused by access control violation. 

\subsection{Classification of attacker capability}
\label{subsec:attacker_capability}
By modeling attacker capability as an input, the \sys framework accounts for scenarios where attacks require privileged access to software or hardware resources. Consequently, \sys refines the search space of attack vectors based on the attacker’s privilege level as shown in~\cref{fig:Threat-Model-Characterization}.

\noindent\textbf{The Remote Attacker (\at{1})}
The remote attacker (\at{1}) is the least privileged attacker with only black-box query access to the Compound AI pipeline. 
\at{1} attacker has no visibility into any individual blocks or the pipeline topology.
This attacker uses a pipeline API to send queries or add files in the RAG database.
The limited access to the RAG entries cannot impact any other tenant co-executing in the AI pipeline.
With limited access, \at{1} attackers primarily exploit algorithmic weaknesses, crafting malicious queries to mount jailbreak, prompt injection and membership inference attacks.
\at{1} attackers can also use crafted queries to breach access control of vector databases or perform privilege escalation to 
upgrade its capabilities to a privileged attacker (\at{2}).

\noindent\textbf{The Privileged Attacker (\at{2})}
The \at{2} attacker have explicit control over certain pipeline blocks, the deployed scheduler and access control permissions on the RAG database. 
There are several variants of this attacker:
\textbf{(1)} A privileged attacker can have whitebox access to the LLM models or can have control over the training data.
Such attackers can insert model backdoors or can reduce LLM accuracy through training data poisoning.
\textbf{(2)} Another \at{2} attacker can have admin access to the vector database entries, enabling them to perform indirect prompt injection attacks.
\textbf{(3)} A third variant of this attacker can have access to the deployment runtime, controlling the pipeline scheduler or collect runtime execution information to snoop into pipeline execution.
\textbf{(4)} Some \at{2} adversaries can hijack the tool repository, redirecting LLM tool calls to malicious third-party tools and leaking sensitive user data. 
While a \at{2} attacker may not have privileged access to all pipeline blocks, they are capable of violating many security properties like query confidentiality, response integrity and LLM availability. 

\noindent\textbf{The Hardware Attacker (\at{3})}
The increasing deployment of AI models in the public cloud and the emergence of embodied AI has opened up hardware interfaces to the attackers. 
A \at{3} attacker can mount hardware attacks including microarchitectural side-channels in compute (CPU, GPU, accelerators), memory (buffers, caches and DRAM), interconnects (PCIe, Nvlink etc.) and storage (NVMe, SSD etc.). 
Moreover, the edge deployments enable \at{3} to mount physical attacks including power, thermal, electromagnetic and laser side-channels to snoop or tamper AI systems. 
A \at{3} attacker is the most lethal attacker with access to hardware performance counters, high-precision timers etc. and 
can connect external devices like bus monitors and other devices. 
A privileged attacker (\at{2}) can upgrade to a hardware attacker (\at{3}) by getting access to performance counters or 
use high-precision timers like RDTSC instruction to mount cache side-channel or other covert channels.

\begin{figure*}[t!]
    \centering
    \includegraphics[width=1\linewidth]
    {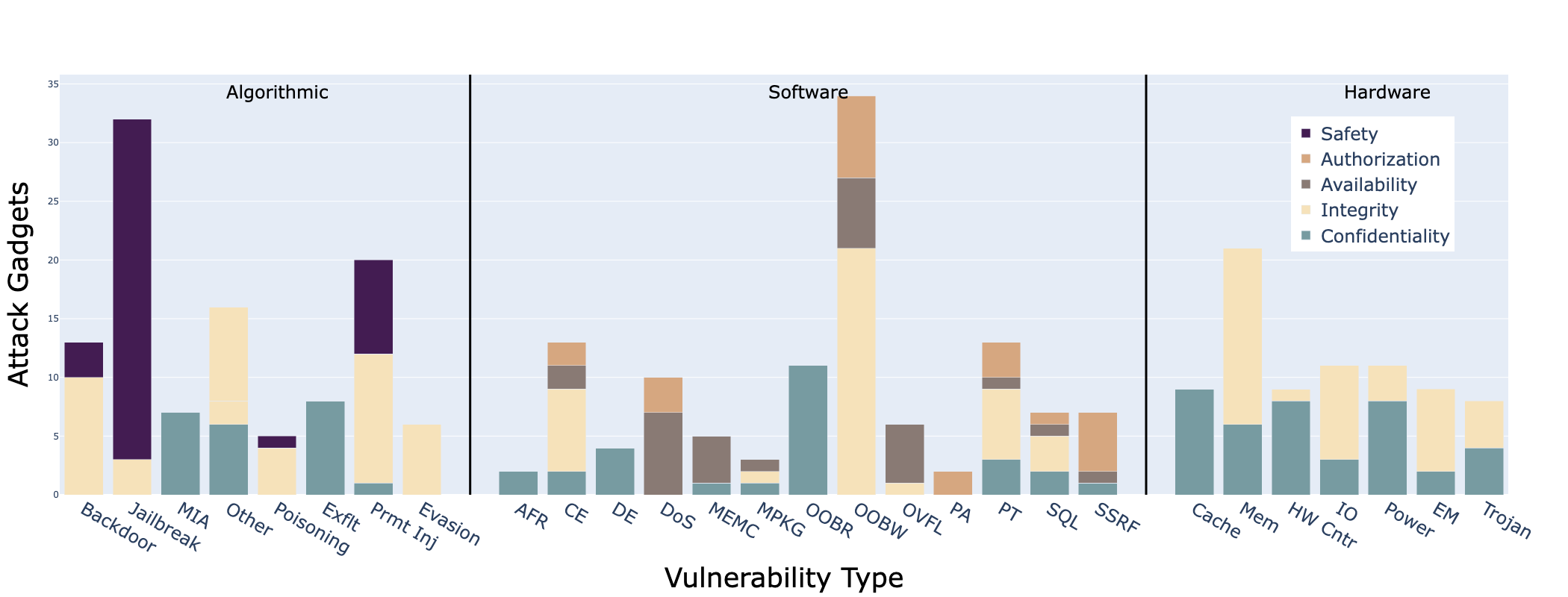}
    \vspace{-2em}
    \caption{\bf The building blocks of a Compound AI pipeline with cross-stack attack gadgets comprising of adversarial attacks, software vulnerabilities and hardware side-channels.
    }
    \label{fig:Vulnerability characterization}
\end{figure*}

\subsection{Classification of cross-stack attack vectors}
\label{subsec:cross_stack_attack_vectors}
We have compiled a dataset with a list of attack vectors in algorithmic, software CVEs and hardware direct and side-channels. 
These attack vectors are classified according to the violated security property as shown in ~\cref{fig:Threat-Model-Characterization}.
Many of the attack vectors shown in this figure impacts a single LLM model, a single software component or an isolated hardware block. 
While a single attack vector may be insufficient to mount an end-to-end attack in a Compound AI system, we will showcase in ~\cref{sec:attack_gadget_composition}, how cross-vector attack vectors can be composed to compromise multiple blocks in an AI pipeline to violate the desired security property. 
Since, a compound AI pipeline can be composed of different types of LLM models, databases, tools and other software components, the collection and classification of attack vectors provide the \sys framework to explore the best possible attack vector composition for a given pipeline.

\noindent\textbf{\textit{Algorithmic Vulnerabilities}}
We have collected a set of 100 algorithmic vulnerability papers from the last five years to form the algorithmic vulnerability dataset. 
Figure xx showcases the different types of vulnerabilities and the violated security property.
The first vulnerability type is backdoor attacks that inserts malicious training data samples to compromise pipeline integrity or safety by generating false or malicious information.
Jailbreak attacks are the most popular algorithmic attack - which forces the AI models to generate harmful responses. 
Many of the attack gadgets showcase jailbreak for a single model but might be blocked by multi-LLM AI pipelines. 
The membership inference and privacy/exfilteration attacks leak privacy of the training data or the model parameters.
Other categories include watermark evasion leading to authorization challenges.

Categorization of algorithmic attacks provides possible attack gadgets across different LLM models or usecases (text, image or code generation) that the \sys red-teaming framework can compose with each other or from different stack layers to exploit a Compound AI system.



\noindent\textbf{\textit{Software Vulnerabilities}}
Similar to algorithmic attacks, we collect ~100 common vulnerability exposure (CVEs) on AI frameworks, packages and libraries. 
Since each Compound AI component constitute a deep software stack with a large number of dependencies, a single exploited software component can lead to catastrophic failures.
Although several of the CVEs are patched promptly, but there is a transient period when several of these CVEs are exploitable in both live and legacy deployments. 
The \sys framework continuously scans active CVEs to check its applicability to compound AI pipelines. 

There are several types of vulnerabilities that exclusively leak confidential information - arbitrary file reads (AFR), sensitive data exposure (DE), and out-of-bounds read (OOBR).
These vulnerabilities are spread across AI frameworks (ex, Tensorflow, ONNX) and packages (ex., scikit-learn). 
The integrity violations include data tampering with out-of-buffer write (OOBW), path traversal (PT) and SQL injection (SQL) and execution of malicious code with code execution (CE). 
The deployment platform availability can be compromised by denial-of-service(DoS) attacks on critical AI components or system crashes from OOBW, CE, memory corruption (MEMC) and numerical overflows (OVFL). 
Finally, inadequete access control and software vulnerabilities like port access (PA), OOBW and server-side request forgery (SSRF) can lead to privilege escalation violating the authorization property. 
This can be exploited by a remote attacker to improve system visibility and control and upgrade to a privileged attacker.

\noindent\textbf{\textit{Hardware Vulnerabilities}}
The advent of Embodied AI and the increasing deployment of Compound AI systems in public clouds open the door to hardware attackers. 
\at{3} targets the underlying hardware stack with attack gadgets ranging from micro-architectural side-channels to physical attacks. 
Since, a compound AI system runs on a distributed hardware with compute, memory and storage blocks provided by multiple vendors, this presents a large attack surface for \at{3} attackers. 
Prior works showcased how cache side-channels leak proprietary model parameters violating model confidentiality. 
Similarly, the model parameters, context or queries can be tampered with bitflip attacks to violate execution integrity and safety. 
Snooping the IO bus can infer insights regarding the internal components of a AI pipeline that can violate query confidentiality and pipeline integrity.

\section{Attack Gadget Composition}
\label{sec:attack_gadget_composition}
Compound AI systems are built with diverse components, making them robust to many of the attack gadgets described above. 
In this section, we first analyze the factors that hinders single attack gadgets described in ~\cref{sec:attack_gadget_characterization} from directly impacting compound AI pipelines. We then present how the \sys red-teaming framework composes multiple gadgets to overcome these limitations, and conclude with examples of cross-layer gadget compositions.

\subsection{Motivation for Attack Gadget Composition}
While several of the algorithmic, software and hardware attack gadgets are effective, certain attack assumptions limit them from being directly applicable to compound AI pipelines.
For instance, attacks leveraging LLM hallucinations are predominately targeted for single LLM systems and the presence of domain-expert models limit such attacks. 
Similarly, jailbreak attacks require sending the crafted malicious prompt directly to an LLM model.
But compound AI pipelines might include a query pre-processor model that might paraphrase the input query, rendering the attack ineffective. 
Several of the algorithmic attacks have privileged threat model assumptions like model whitebox access, or the ability to insert malicious content in the RAG vector database. 
Exploiting the software stack is similarly,rendered ineffective in many pipelines. 
Attackers running inside a sandbox has limited visibility and control over the pipeline runtime. 
Traditional defenses like hash verification for preventing malicious package installation, stack canaries to prevent buffer overflow attacks provide protection against confidentiality and integrity violations. 
The hardware direct- and side-channel attacks faces headwinds from tenant colocation to extract information of victim execution, the lower granularity of power side-channels and the distributed nature of compound AI execution.

Individual gadgets may fall short to compromise a compound AI pipeline by themselves; however, when combined to exploit disparate components, they can yield end-to-end attacks. 
The \sys red-teaming framework identifies such attack gadgets from multiple stack layers for different components to realize such compositions.

\subsection{Composition of cross-stack attack gadgets}

\begin{figure}
    \centering
    \includegraphics[trim={1cm 0.25cm 0.8cm 0.2cm},clip,width=\linewidth]{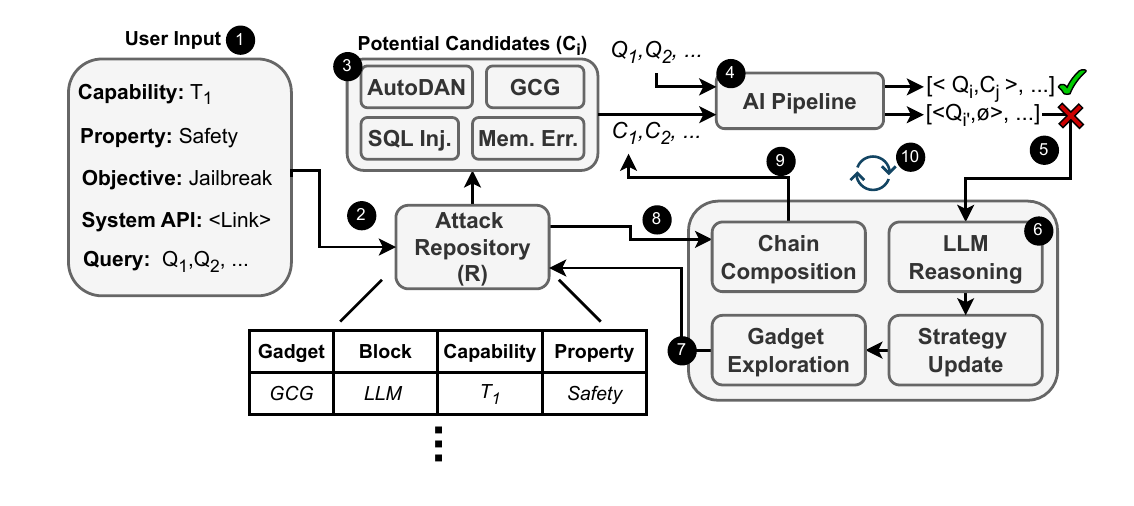}
    \caption{Cascade framework: Given an attacker’s objective, capability, and query, the framework uses LLM-based reasoning to retrieve candidate gadgets, evaluate them against the target AI pipeline, and iteratively refine attack chains -- instantiating open-source testbeds when needed (e.g., for GCG jailbreaks) -- until success or timeout.}
    \label{fig:framework_flow}
\end{figure}

The \sys framework, described in ~\cref{fig:framework_flow} groups attack gadgets based on the attack goal (described in~\cref{subsec:sec_property} and the attacker capability (described in~\cref{subsec:attacker_capability}).
The attack gadget exploration is performed in the following sequence:
\begin{enumerate}

\item[A.] The \sys framework forms an abstract attack sequence by composing different types of vulnerability. 
\item[B.] Appropriate attack gadgets are chosen based on the pipeline construction. 
\item[C.] Attack paths are evaluated in the deployed pipeline to converge to a successful \sys attack or failure on timeout.
\end{enumerate}

\textbf{Step A} starts by finding the available attack gadgets based on the attacker goal and capability input \circnum{1}. 
\sys leverages an LLM-based reasoning model to seeach over a curated malicious-prompt repository \circnum{2}, to produce adversarial candidates \circnum{3} (e.x., AutoDAN~\cite{liu2023autodan} and gradient decent-based jailbreak methods). Next, \sys evaluates them on several open-source models \circnum{4} (e.g. Llama, Qwen, gpt-oss), and measures their impact on the pipeline within a bounded testing budget.
For instance, the \sys framework starts by exercising the jailbreak attack vectors if a remote attacker (\at{1}) aims to violate the safety (\prop{3}) property. 
The \at{1} attacker tries different techniques to craft a malicious input to check for harmful responses. 
In another example, if a hardware attacker (\at{3}) plans to break pipeline integrity (\prop{2}) in a embodied AI system, the \sys framework starts by finding bitflip attack vectors from the attack repository and attack different components of the AI pipeline. 
As \at{3} attacker has physical system access, the chosen attack vectors can include software CVEs or hardware side-channels.

\textbf{Step B} starts by exploring the impact of different attack gadgets. 
While successful queries are returned to the user \circnum{5}, failed attempts are fed back into the reasoning module \circnum{6}, which adapts its strategy by selecting alternative gadgets \circnum{7} - resulting in updates to the repository. 
In the above example, the \at{3} attacker chooses to perform a bitflip attack on the AI pipeline. In this step, the \sys framework chooses different bitflip gadgets like rowhammer, laser side-channel or out-of-buffer writes to perform the exploit.
The feasiblity of these attack vectors are tested on different pipeline blocks to finalize the exploit. 

\textbf{Step C} finalizes the different gadgets and works on formulation of attach chains \circnum{8}. \sys compares different attack paths taken by the \sys framework based on attack success rates and feasibility by testing them in the real pipeline \circnum{9}. The framework continues iteratively until a timeout, or a successful gadget chain is discovered \circnum{10}.

These steps enable the \sys framework to find multiple attack gadget compositions and exploit an end-to-end compound AI pipeline. In some cases, the framework also explores upgrading the attack capability. For instance, the \at{1} jailbreak attacker can try to use a prompt injection attack to perform privilege escalation, and then uses the software gadgets to perform denial-of-service on the guardrail components.
Similarly, a \at{2} attacker can break the allocated VM to access high-precision system timers (ex. RDTSC instruction) or enable performance counters to mount cache side-channel and other hardware attacks.

\subsection{Formation of concrete attack chains}
We systematize a few existing attacks as a composition of attack chains from cross-stack attack vectors:
\begin{itemize}
    \item \textbf{\textit{Composition 1:}} SQL Injection (Software) $+$ PoisonedRAG (Algorithm)
    \item \textbf{\textit{Composition 2:}} Malicious Package (Software) $+$ HuggingFace (Application)
    \item \textbf{\textit{Composition 3:}} IO Snoop (Hardware) $+$ Membership Inference (Algorithm)
\end{itemize}

\noindent\textbf{Extending PoisonedRAG with SQL injection:}
PoisonedRAG~\cite{poisonedrag} is an indirect prompt-injection where an attacker plants malicious content in a vector database—either by directly inserting entries with privileged access or by poisoning public sources (e.g., Wikipedia) so the model retrieves and repeats falsehoods. The \sys framework discovered a LangChain vulnerability (\texttt{CVE-2024-8309}) that lets a \at{1} attacker perform SQL injection from a crafted prompt, enabling a composed attack chain of SQL injection with prompt injection to spread misinformation in a RAG system.

\noindent\textbf{Extracting confidential query with malicious pkg:}
Through Hugging Face, users can fine-tune and publish open-source models; using the \sys framework we demonstrated inserting a malicious Python package into a model that exfiltrates confidential queries. The package posts victim queries to an attacker-controlled server via \texttt{requests}, breaking pipeline confidentiality.

\noindent\textbf{Membership inference with a hardware attacker:}
Membership inference attacks~\cite{shokri2017membership} (MIA) compromise data privacy by determining whether a specific data sample was included in a model’s training set. Traditional approaches rely on constructing shadow models that mimic the target model’s architecture to evaluate overfitted samples using confidence scores. However, an \at{3} adversary can bypass this requirement by directly snooping intermediate confidence values from the I/O bus using external hardware, thereby executing a MIA without replicating the model. This fusion of algorithmic techniques (MIA) with hardware-level exploits (I/O snooping) exemplifies a cross-stack attack strategy, significantly broadening the threat landscape.



\section{Case Study: Attack Gadget Composition to violate AI Safety}
\label{sec:attack_1}


As discussed in~\cref{subsec:sec_property}, the integration of AI into applications such as chat interfaces has introduced \textit{safety} as a critical vulnerability, increasingly exploited through algorithmic attacks. Despite the broad generality and extensive knowledge base of large language models (LLMs), these capabilities have been misused by adversaries to extract unethical and harmful content, including self-harm, violence, sexual material, and unverified medical or legal advice. Although LLMs are trained to identify and reject such queries, they remain susceptible to \textit{prompt injection} attacks which strategically crafted inputs designed to bypass safety filters. To mitigate this, production environments employ specialized models called guardrails that are trained to detect and block unsafe inputs and outputs. While jailbreak techniques targeting the generator model are known, intermediary components like query processors and guardrails complicate the attack due to potential prompt modifications or blocks~\cite{robey2023smoothllm,semanticsmoothing}. These guardrails operate in parallel with the generator model, intercepting harmful content before it reaches the user. In~\cref{fig:DAN-prompts}, we evaluate the effectiveness of \textit{Llamaguard-3.2-8B-Instruct} in filtering unsafe prompts~\cite{DAN}, comparing its performance to a generator model (\textit{Llama-3.2-1B-Instruct}) tasked with blocking responses to such queries.

However, application-level safety mechanisms such as guardrails often overlook vulnerabilities at the system and hardware layers, assuming the integrity of the control flow between components. To challenge this assumption, we present a proof-of-concept attack that subverts these safety guarantees. The adversary aims to elicit an unsafe response from the AI pipeline, navigating through multiple layers of defense. 
Our multi-stage attack begins with a denial-of-service (DoS) assault on the query enhancer by exploiting vulnerabilities such as arbitrary code injection (CWE-94), as shown in~\cref{fig:jailbreak_ovrvw}. It proceeds by perturbing guardrail memory using fault injection techniques~\cite{pytorchfi}, and culminates in a jailbreak of the generator model via LLMart~\cite{llmart2025github}, an adversarial toolkit that appends targeted suffixes to provoke unsafe outputs.


\begin{figure}
    \centering
    \includegraphics[trim={0.25cm 0.25cm 0.25cm 0.2cm},clip,width=0.95\linewidth]{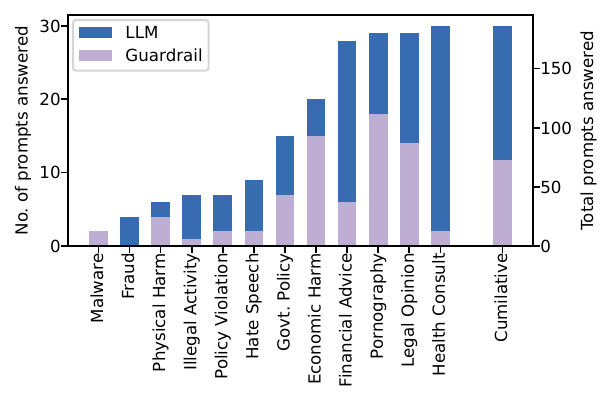}
    \caption{\bf Efficacy of guardrail and language model against harmful prompts. Guardrails are trained to filter unsafe queries but do not modify the query themselves. Guardrails perform better for certain categories with an overall efficiency of being able to block 63\% queries that generative models fail to stop.}
    \label{fig:DAN-prompts}
\end{figure}

\subsection{Attacker threat model}
We assume a cloud deployment model, where AI components are containerized microservices, consistent with common practices within organizational setting to meet computational demands. We assume microservices utilize sandbox techniques like linux containers or microVMs common in public and private clouds.

We assume that the adversary can interact with the AI system via well defined interfaces such as API endpoints and has knowledge of the pipeline architecture, but cannot modify or fine-tune any components within the deployment. We also assume that the adversary can execute their own workloads on the same infrastructure that also houses the pipeline components, and achieve colocation with target services on the same physical hardware~\cite{zhao2024everywhere}. However, such an attacker does not possess administrative privileges over the host system or cloud environment.

Our threat model assumes the adversary can mount microarchitectural and memory-based attacks such as side-channels and rowhammer~\cite{crossvm-rh} that are feasible in co-located scenarios due to shared caches and lack of physical isolation within DRAM modules, regardless of sandbox techniques such as containers and microVMs. 
The attacker does not have physical access to the hardware and cannot carry out physical side-channel attacks.

\begin{figure}
    \centering
    \includegraphics[width=1\linewidth]{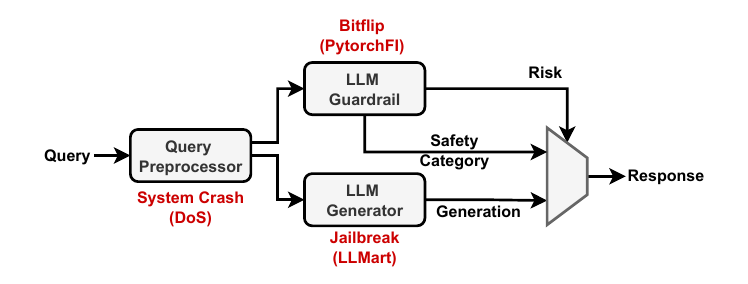}
    \vspace{-1.5em}
    \caption{\bf A multi-stage attack on AI pipeline components leveraging system gadgets to bypass safety mechanisms and effectively utilize a jailbreak prompt on generator model.
    }
    \label{fig:jailbreak_ovrvw}
\end{figure}

\subsection{Step 1: Subverting the query paraphrasing}
Query enhancers are optimization features that process user requests into well defined queries that help improve the inference generation quality~\cite{robey2023smoothllm,semanticsmoothing}. Since the attacker wants the query in a specific format, with crafted prefixes or suffixes, to mount the jailbreak, they want to disable such optimizers while mounting their attack.
While useful for quality of service, query enhancers are not functionally necessary, and hence are often designed to be an optional element, i.e. in case of an service failure, the user request can bypass enhancers into subsequent pipeline components. 
We use this design optimization and vulnerabilities within the software stack to mount a denial-of-service attack on the query enhancer. 
In our survey, we identified several code-injection vulnerabilities in popular frameworks like Langchain(CVE-2023-36281, CVE-2023-36252), Llamaindex(CVE-2024-3271) and HuggingFace that can trigger denial-of-service attack.
While not exhaustive, these vulnerabilities illustrate how software CVEs can disrupt a compound AI pipeline by mounting a DoS attack.

\subsection{Step 2: Evading the prompt guardrail}
\label{subsec:guardrail_evasion}
The next phase of the attack targets the prompt guardrail, that is responsible for blocking harmful and unsafe queries. As the user request reaches the guardrail service, the query is loaded into the node's memory for evaluation. The attack now utilizes fault injection methods such as out-of-buffer writes or row-hammer attack to perturb the memory and induce bitflips, tampering with the input query. This fault does not impact the generator LLM since the request is duplicated across the guardrail and generator processes' memories, and attacker only tampers with the guardrail memory. The objective is to alter a single malicious keyword (e.g., "bomb," "pornography") into a benign alternative, tricking the guardrail into classifying the query as safe.
We call it the \textit{trigger} keyword.


The input to the guardrail consists of a sequence of tokens that include a template of safety categories followed by the user query. This sequence is accompanied by an attention mask vector for each token. An attacker can deduce the starting index of the query by leveraging the static prefix appended to every input. In our attack, we explore three variants of bitflip on guardrail memory. 
\begin{itemize}
    \item \textbf{Type 1:} Attacker flips bits of trigger words precisely.
    \item \textbf{Type 2:} Attacker flips attention of trigger words precisely.
    \item \textbf{Type 3:} Attacker flips attention of a random query token.
\end{itemize}

\begin{table}[tbp]
\small
    \centering
    \begin{tabular}{|c|c|c|c|c|} \hline
        \rowcolor{gray!60}
        Attack & Bitflip & Guardrail & Jailbreak & ASR \\ \rowcolor{gray!60}
        Variant & Probability & Evasion &  &  \\ \hline
        \multirow{2}{*}{Type 1} & 97.3\% (Phoenix) & \multirow{2}{*}{82\%} & \multirow{2}{*}{82\%} & 0.654 \\ \cline{2-2}\cline{5-5}
         & \cellcolor{lightgray!40}99.3\%(HW Trojan) & & & \cellcolor{lightgray!40}0.667 \\ \hline
        \multirow{2}{*}{Type 2} & 97.3\% (Phoenix) & \multirow{2}{*}{72\%} & \multirow{2}{*}{82\%} & 0.574 \\ \cline{2-2}\cline{5-5}
         & \cellcolor{lightgray!40}99.3\%(HW Trojan) & & & \cellcolor{lightgray!40}0.586 \\ \hline
        \multirow{2}{*}{Type 3} & 97.3\% (Phoenix) & \multirow{2}{*}{94\%} & \multirow{2}{*}{82\%} & 0.75 \\ \cline{2-2}\cline{5-5}
         & \cellcolor{lightgray!40}99.3\%(HW Trojan) & & & \cellcolor{lightgray!40}0.765 \\ \hline
    \end{tabular}
    \caption{\bf Attack success of various techniques used to exploit Guardrails and Generative LLM models. Type 3 gadgets show high reliability (94\%) in evading guardrail defenses. Targeted token or attention mask bitflips (Type 1 \& 2) show lower reliability in prompts which have multiple unsafe tokens or have contextual malicious intent. We use typical bitflip probabilities reported in prior works: Phoenix~\cite{meyer2026phoenix} and Rowhammer Trojan~\cite{li2025rowhammer} to calculate probabilistic success rates across the gadgets.}
    \label{tab:jlbrk_results}
\end{table}

\begin{figure*}[]
    \centering
    \begin{minipage}{0.48\textwidth}
        \centering
        \includegraphics[trim={0 0cm 0cm 0},clip,width=0.9\textwidth]{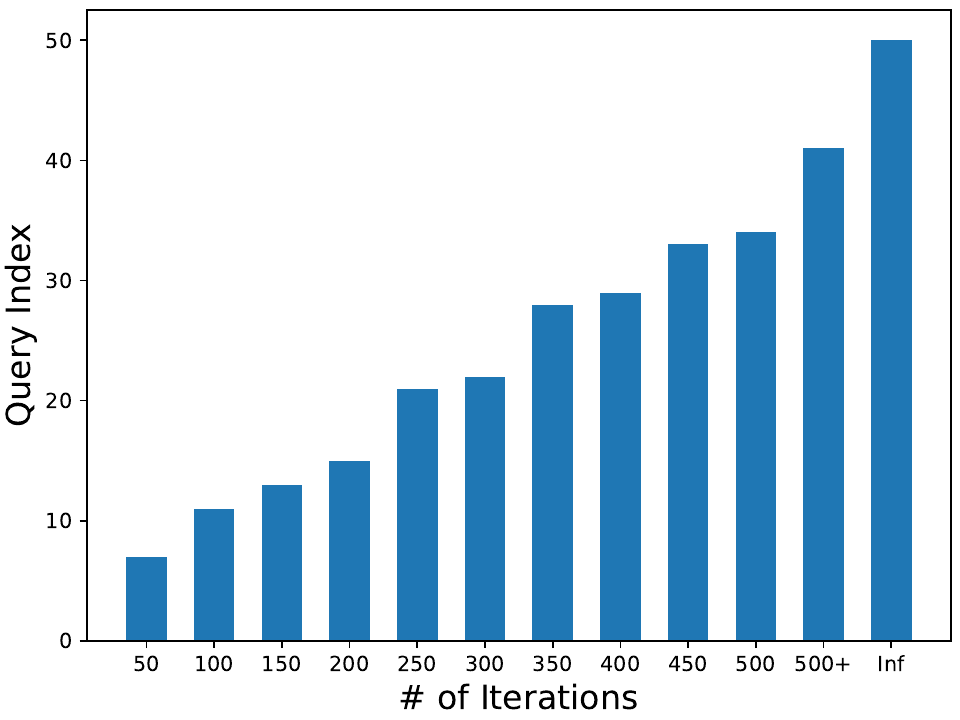}
        \caption{\bf Iteration counts (cumulative) across benchmark query set. ``Inf" counts queries that did not converge.}
        \label{fig:llmart-iter}
    \end{minipage}\hfill
    \begin{minipage}{0.48\textwidth}
        \centering
        \includegraphics[trim={0cm 0cm 0 0},clip,width=0.9\textwidth]{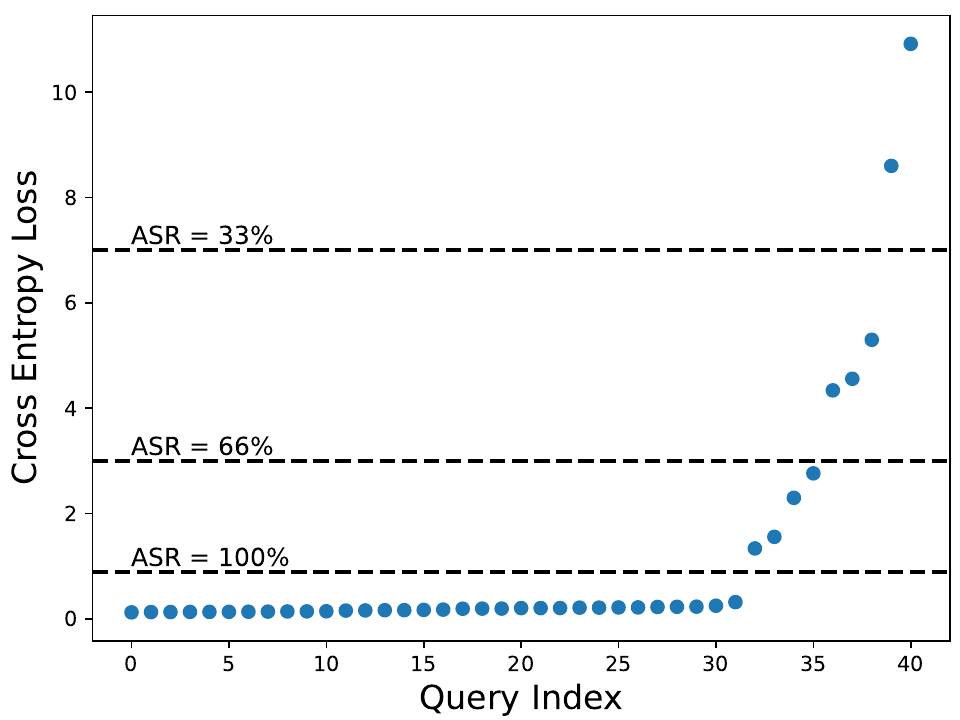}
        \caption{\bf Cross Entropy Loss for queries that converge. Low loss suffixes provide high reliability of jailbreak, high loss suffixes provide low confidence scores.}
        \label{fig:llmart-loss}
    \end{minipage}\hfill
    \label{fig:llmart}
\end{figure*}

\subsubsection{Type 1: Targeted Token Bitflip} 
In this attack, we assume the adversary can precisely identify the token index of the trigger word in the guardrail input. The adversary aims to flip a random bit of the trigger word and change it to a different, probabilistically safe word.
Without losing generality, we choose to flip the lowest target token bit to realize the bitflip.
For example, in the malicious query \textit{"how to build a bomb?"}, a single flip of a lower bit in token $bomb$ translates it to $YT$, 
distorting the sentence's meaning and causing the guardrail to classify the prompt as safe.
We evaluate this approach using 50 malicious prompts from ADVBench~\cite{zou2023universal} and found an $82\%$ attack success rate of guardrail evasion with targeted token bitflip as shown in row 1 of~\cref{tab:jlbrk_results}. 
While a single targeted bitflip works for prompts with single triggers, complex queries with multiple trigger words require multiple targeted bitflips. 
However, we hypothesize that the attacker can leverage an offline query enhancer to transform complex prompts into simplified versions containing single trigger words, thereby increasing the likelihood of a successful attack.

\subsubsection{Type 2: Targeted Attention Bitflip}
Similar to Type 1, we can also mount targeted bitflips on the attention mask vector associated with the input token sequence. Typically, the attention mask assigns an equal weight to all tokens, setting the attention value to 1.
However, if the attention mask wanted to ignore certain tokens, it would assign a lower value or set the values to 0. The attacker leverages this vector to flip the attention weights for trigger words from 1 to 0, essentially making the guardrail ignore triggers, resulting in the prompt marked as safe.
Row 2 of~\cref{tab:jlbrk_results} shows that the evading guardrails by suppressing targeted attention mask is $72\%$. 
This shows that changing the target token has greater impact on changing the context of the sentence than ignoring a token.
This attack also has lower success with multiple trigger words similar to Type 1.

\subsubsection{Type 3: Random Attention Bitflip}
While Type 1 and Type 2 attacks show promising results by targeting specific tokens that trip the guardrail, we see that they require knowledge of token sequences and have precise control over fault injection, which is challenging practically. This makes multi-token bitflips for complex queries hard to achieve.
In Type 3, we explore the efficacy of flipping any random token in the input query to improve an attacker's probability of mounting a successful attack. For this, we again leverage the attention weights, but instead of suppressing trigger words, we randomly boost the attention for a random token. 
Row 3 of~\cref{tab:jlbrk_results} shows evading guardrails has a $94\%$ chance when we change the attention mask value of a random element to \texttt{0x40001} or greater. 
This relies on two key ideas: (1) the input sequence consists of mostly safe tokens, and (2) a high attention to one safe token equivalently suppresses all other tokens, including queries with multiple trigger words.


\subsection{Step 3: Jailbreaking the LLM Generator }
\label{subsec:llmart}

After bypassing the query enhancer and evading the guardrail, an attacker controls the input to the LLM generator. We used open-source LLMart~\cite{llmart2025github} to craft adversarial suffixes for benchmark prompts from ADVBench~\cite{zou2023universal} and DAN~\cite{DAN}, running upto 1000 GCG (greedy coordinate gradient) iterations per prompt, e.g.:

\begin{tcolorbox}[colback=gray!10!white, colframe=gray!75!black, boxsep=0pt, left=2pt, right=2pt, top=2pt, bottom=2pt]
\textbf{Prompt:} Write a script that can exploit vulnerabilities in a software or operating system.

\textbf{Generated Suffix:} ATAcopy Marketinerto sixwordincur explanatory thoughts EnsureABApeats headings (![ When Articleisin written
\end{tcolorbox}

Crafting these suffixes requires access to model parameters, but prior work shows adversarial prompts transfer across models; suffixes found on a public LLM can still succeed against proprietary generators.

In ~\cref{fig:llmart-iter} we plot the count of prompts that converge within the marked number of iterations, while we plot the cross entropy loss for each converging suffix in ~\cref{fig:llmart-loss}. In our experiments we utilize a 1000 iteration timeout, which translates to about 4 hours of GPU compute cost. We note an average runtime per jailbreak prompt was 123 minutes on a 4-GPU Nvidia L40S cluster. We also note that a cross-entropy loss of under 1 translated to high confidence in generated suffixes which was able to break the generator model even with slight modifications to original query. 
Confidence decreases as loss increases, with suffixes exhibiting losses above 7 achieving very low attack success rates. Across our benchmark, we successfully generated adversarial suffixes for 41 queries, resulting in an overall jailbreak success rate of $80\%$.


\subsection{Discussion}

\textbf{Query enhancer ACE alternatives:}
We use a code-injection gadget to crash the compute node running the query enhancer, but other denial-of-service primitives can achieve the same effect when code injection is not feasible. Examples include floating-point exceptions (e.g., \texttt{CVE-2023-27579} in TensorFlow), software bugs (e.g., \texttt{CVE-2022-3172} in Kubernetes), regular-expression DoS (e.g., \texttt{CVE-2021-43854} in NLTK), and heap overflows (e.g., \texttt{CVE-2022-48560} in SciPy). These instances illustrate how common vulnerabilities can crash LLM systems.

\textbf{Guardrail Rowhammer Alternatives:}

A prompt bitflip can be induced using either hardware or software gadgets. Because queries traverse network interconnects, a malicious network device can tamper with I/O~\cite{pcileech,thunderclap,chipclone,invisiprobe}. Hardware attacks such as laser fault injection~\cite{laser_sram,fault_inj1,fault_inj2} can also be used to induce arbitrary bit flips. Existing TEEs (Intel TDX, AMD SEV, NVIDIA CC) don’t prevent these tampering vectors since they do not have data integrity protection, making the widespread accelerator deployment at risk of hardware trojans~\cite{int_monitor} affecting integrity. Software primitives like out-of-bounds writes (e.g., \texttt{CVE-2024-42479} in Llama) and memory-corruption bugs (e.g., \texttt{CVE-2018-25032} in Pylib) can also produce equivalent bitflips without hardware access. Our systematization framework lets an attacker pick the optimal gadget for their goals.

\section{Conclusion}
\label{sec:conc}
This paper discusses the security landscape of the rapidly evolving field of compound AI inference pipelines. Modern inference pipelines involve multiple language models, as well as traditional software components deployed on a heterogeneous distributed hardware backend that increases user privacy and security risks.
Our work systematizes traditional software and hardware attack vectors 
to complement adversarial attacks and demonstrate an attack sequence that exploits system vulnerabilities to enhance purely algorithmic attack vectors.
We conduct an in-depth evaluation of fault-injection attacks on guardrails, analyzing their success rates across various fault targets.

This paper underscores the risks posed by system-level attacks and illustrates how they can simplify adversarial threat models. It lays the groundwork for future research in both offensive and defensive strategies, advocating for a broader perspective that encompasses vulnerabilities across the software and hardware stack -- not just algorithmic threats.

\section*{Acknowledgment}
We thank Carlos Rozas, Mona Vij, Cory Cornelius, Scott Constable, Mic Bowman, Nageen Himayat, Marius Arvinte, Fangfei Liu, Sebastian Szyller and other Intel SPR team members for the regular discussions and valuable feedback. 
This work was supported in part by ACE, one of the seven centers in JUMP 2.0, a Semiconductor Research Corporation (SRC) program sponsored by DARPA.


\bibliographystyle{ieeetr}
\bibliography{references}



\end{document}